\documentclass{article}
\usepackage{graphicx} 
\usepackage{amsmath}
\usepackage[numbers]{natbib}
\usepackage{multirow}
\usepackage{threeparttable}
\usepackage{capt-of}
\usepackage{rotating}
\def\boxit#1{%
  \smash{\fboxrule=1pt\relax\fboxsep=2pt\relax%
  \llap{\rlap{\fbox{\vphantom{0}\makebox[#1]{}}}~}}\ignorespaces
}

\begin{document}
\begin{center}
\Large \textbf {A Roadmap of Emerging Trends Discovery in Hydrology: A Topic Modeling Approach}\\
\end{center}
\author{Sila Ovgu Korkut$^{1*}$, {O}znur Oztunc-Kaymak$^2$, Aytug Onan$^3$,\\ Erman Ulker$^4$, and Femin Yalcin$^1$ }
\begin{center}
\hspace{-2 cm}$^1$ Dept. of Engineering Sciences, Izmir Katip Celebi University, Izmir, 35640, Turkey\\
silaovgu.korkut@ikcu.edu.tr, femin.yalcin@ikcu.edu.tr\\
\vspace{0.1 cm}
\hspace{-2 cm}$^2$ Izmir Democracy University, Information Technology Department, Izmir, 35140, Turkey.
oznur.oztunckaymak@idu.edu.tr\\
\vspace{0.1 cm}
\hspace{-2 cm}$^3$ Dept. of Computer Engineering, Izmir Katip Celebi University, Izmir, 35640, Turkey.
aytug.onan@ikcu.edu.tr\\
\vspace{0.1 cm}
\hspace{-2 cm}$^4$ Dept. of Civil Engineering, Izmir Katip Celebi University, Izmir, 35640, Turkey.
erman.ulker@ikcu.edu.tr\\
\end{center}
\hspace{-1 cm} \textbf{Corresponding author: Sila Ovgu Korkut$^{*}$ (silaovgu.korkut@ikcu.edu.tr)}
\begin{abstract}
In the new global era, determining trends can play an important role in guiding researchers, scientists, and agencies. The main faced challenge is to track the emerging topics among the stacked publications. Therefore, any study done to propose the trend topics in a field to foresee upcoming subjects is crucial. In the current study, the trend topics in the field of "Hydrology" have been attempted to evaluate. To do so, the model is composed of three key components: a gathering of data, preprocessing of the article's significant features, and determining trend topics. Various topic models including Latent Dirichlet Allocation (LDA), Non-negative Matrix Factorization (NMF), and Latent Semantic Analysis (LSA) have been implemented. Comparing the obtained results with respect to the $C_V$ coherence score, in 2022, the topics of "Climate change", "River basin", "Water management", "Natural hazards/erosion", and "Hydrologic cycle" have been obtained. According to a further analysis, it is shown that these topics keep their impact on the field in 2023, as well.
\end{abstract}

\section{Introduction}
\label{intro}
In the new global age, academic research has a vital role in the enhancement of technology and the creation of development plans for countries. Academic research also sheds light on future plans, problems or measures to be taken, and investments. Therefore, identifying the main topics, that have been pointed out in recent years to make decisions about what to do in the next few years, is crucial. Due to the rapid progress of technology, the volume of data grows in literature. This implies some issues such as keeping track of new research, controlling the novelty of a paper, and exploring a new topic for new research, \cite{schaefermeieretal,zhangetal}. Moreover, the increase in the number of interdisciplinary studies and more holistic approaches in scientific papers are another cumbersome for researchers who need qualitative information. Taking into account the ”publish-or-perish principle” \cite{vanDalen}, any assistance and development in the qualitative determination of research topics gain significant importance to evaluate research trends, emerging fields, and research frontiers. That’s why any automated contribution to the research community to manage huge data for decision-making is more attractive. The main motivation of this study is to explore the trend topics in the field of "Hydrology" and to highlight the keywords on where the studies conducted in recent years have evolved for interested readers.

Hydrology is the science of water that deals with mechanisms of precipitation, evaporation, transpiration, and infiltration of water states at the hydrosphere. These four mechanisms are evaluated together in a loop, which is called the hydrologic cycle. Variability in any mechanism has a direct effect on another, therefore, it brings complexity to bring back to balance in order to avoid any climatic changes. The broad scope, rapidly changing interests, a fast transition of research topics in hydrology, and increasing climate change effects make hydrology an interesting field for trend studies.

The current study attempts to illuminate the emerging topics by employing three topic models: the Latent Dirichlet Allocation (LDA), the Non-negative Matrix Factorization (NMF), and the Latent Semantic Analysis (LSA). The difference between our scientific research approach and the classic trend analysis studies found in the literature is that it offers researchers in the field of real hydrology insight. We believe that the results obtained with this approach will be a source of inspiration for researchers by presenting the trending issues in the field of ”Hydrology". Therefore, the two primary scopes of the current study can be summarized: First to compare the results of the well-known topic models in the field of hydrology, and second to determine the emerging topics in the field of "Hydrology" and present recommendations to the literature by evaluating the topics in line with opinions. To the best of the authors' knowledge, this is the first effort to do such a study utilizing topic models in the field of hydrology.

The paper is composed of four themed sections. It begins first by overviewing the related works in the literature. Section~\ref{sec:method} begins by laying out the theoretical dimensions of the research and presents the details of the methodologies. Then, the attained results and discussions of the research are presented in Section~\ref{sec:results} focusing on the three key themes: comparison of the topic models, interpreting and testing the preferred model results, and providing future projections. Finally, Section~\ref{sec:con} outlines the main key points of the study.

\section{Related Works}
\label{sec:relatedwork}
Recent evidence suggests that investigating the identification of existing and emerging research trends is having an increasing impact on the scientific literature. There are various methodologies that attempt to explore emerging topics in the literature. Chen, \cite{Chen}, aimed to present theoretical and methodological contributions utilizing a time-variant duality between research fronts and intellectual bases where research fronts stand for evolving concepts of research issues and the intellectual base is associated with the research's citation and co-citation footprint in the scientific literature. On the other hand, Jiang et al. \cite{jiangetal} proposed a methodology that determines key phrases and authoritative authors by ranking authors and phrase information via RNN.

While the above-mentioned papers focused on the variety of attributes for modeling trends, we have been witnessing the increased impact of the use of topic models for diverse fields since the rise of topic models. For instance, a bibliometric-based topic modeling approach has been used in \cite{JIANG2016226} to determine the hot spots of hydropower research. The conducted study took 1726 articles into account covering 19 years time span from 1994 to 2013 and returned 29 topics in the specified keyword. Alnusyan et al. \cite{Alnusyan} proposed a semi-supervised approach to the learning of latent topics over reviews on the Amazon website. A systematic literature review has been conducted in \cite{KIM2022263} to recommend the future research direction in humanitarian relief logistics (HRL). 1058 papers have been analyzed therein by using the concepts of keyword network analysis, keyword frequency analysis, network analysis, and topic modeling. Moreover, five different topic models, that are Latent Semantic Analysis, Latent Dirichlet Allocation, Non-negative Matrix Factorization, Random Projection, and Principal Component Analysis, have been studied in \cite{albalawietal} considering user-generated content. The performances of the identified models have been compared by considering recall, precision, F-score, and topic coherence metrics. The topic space trajectories analysis method has been presented in \cite{schaefermeieretal} for tracking research topics. The method has been tested in machine learning literature within a 50-year span. Moreover, the materials-aware language model called MatSciBERT which is mainly based on Bidirectional Encoder Representations from Transformers (BERT) models has been proposed in \cite{Guptaetal}. The model has been trained on a large corpus of peer-reviewed materials science publications. A survey of literature analysis tasks has been presented based on both word representation learning-based models and the graph representation learning-based models in \cite{zhangChen}. Furthermore, in more recent years, four topic models, which are Latent Dirichlet Allocation (LDA), Non-negative Matrix Factorization (NMF), Top2Vec, and BERTopic, have been discussed in the study of \cite{egger_yu} by taking Twitter posts as the reference point. Moreover, 51346 abstracts from 23 prestigious journals in structural engineering with a publication period from 2000 to 2020 have been analyzed in \cite{XIE2022577} using the Latent Dirichlet Allocation (LDA) to extract the 50 emerging topics.
\section{Data Source and Methodology}
\label{sec:method}
The main goal of this section is to present details of the components of our trend detection model, rigorously. The model is formed by three essential components: a collection of data, boosting the key features as well as the body of the article, and extracting the points of prediction.

\subsection{Data Collection and Analysis}
\label{sec:data_collection}
The Web of Science Core Collection (WOS) \cite{journalindex} is one of the most common databases to fetch bibliographic information for literature research \cite{guz}. For many years, WOS by Thomson Reuters (ISI) was the only citation database and publication covering all areas of science \cite{agha}. One of the strongest aspects of the WOS database is that it has citation and bibliographic data dating back to the 1900s \cite{boyle}. Due to these capabilities, it is even claimed to be the deepest and highest quality \cite{agha}. Therefore, as many as 56951 articles' bibliographic information on the "Hydrology" keyword has been collected from the WOS database for this study. The accomplished analysis has been conducted for the period of 01 January 2008 to 24 March 2022. The shape of the initial dataset is composed of articles, conference proceedings, books, book chapters, letters, reviews, biographical items, and editorial notes. However, the final dataset contains 51776 articles after conducting a careful analysis of the data set. To do so, articles and proceeding papers were analyzed in this study as research items on the basis of the consultant of the experts. Before dropping the null values, several manipulations were attempted to manipulate the data. For instance, taking the high correlation between author keywords and the keywords assigned by WOS to the research item into account, the null values were filled by combining these two features. Besides, the publication year, which can have an important role in the trend, had null values which of those items had the early access date values. Therefore, the null values of the publication year were assigned by using the year of early access date. The process of data preparation was completed by removing the null values which cannot be manipulated.

\subsection{Feature Extraction of the Textual Data}
\label{sec:text_mining}
Existing research recognizes the critical role played by text mining has become a central issue for research trends.
The process for the data analysis has been followed by textual analysis. For that purpose, "Article Title", "Keywords", and "Abstract" are centered as the eligible features of an article. In addition to these, WOS categories have been taken into account. Based on the scope of the current study, four subsets constructed from possible combinations were designed to conduct the comparative work. For a comprehensive analysis, three topic models were selected: Latent Dirichlet Allocation, Non-negative Matrix Factorization, and Latent Semantic Analysis.

The text pre-processing steps followed the lines of standard procedure as follows: tokenization, stemming, lemmatization, noise and stopwords removal, and filtering tags. To do so, the Natural Language Toolkit (NLTK) \cite{hardeniya2016natural} has been utilized. After completing the standard procedure, the verbs and their possible conjugations as well as the stem of the "hydrology" are removed to eliminate the dominance. The uni-gram phrase model is analyzed via the use of the \verb|sklearn| library, \cite{scikit-learn}.

\begin{figure}[ht!]
\centering
\includegraphics[scale= 0.42]{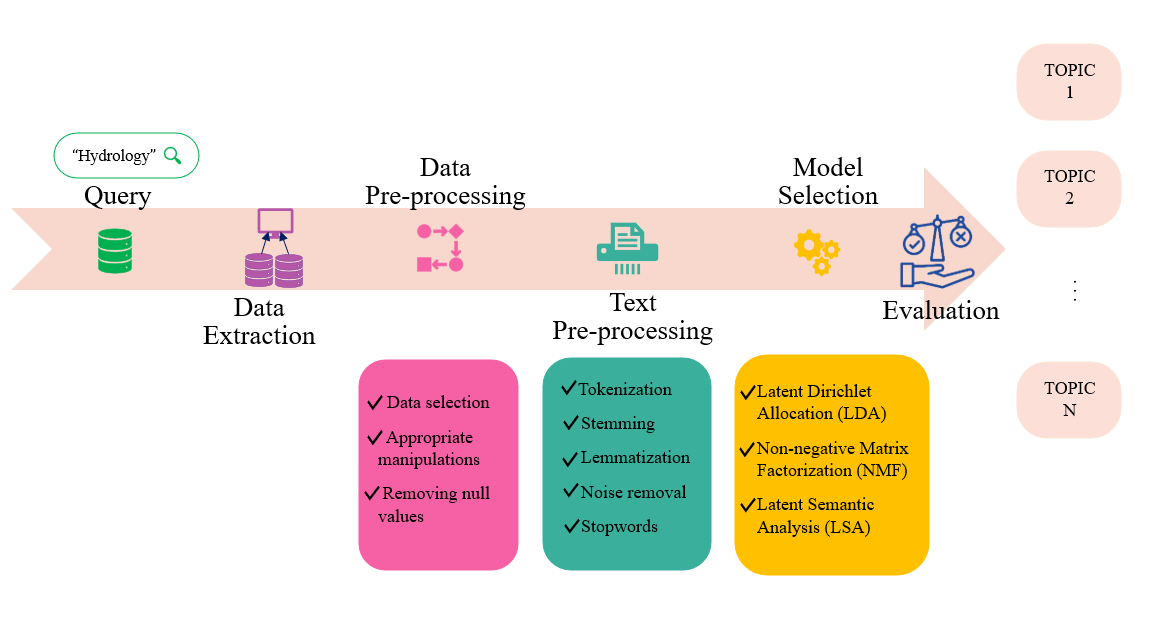}
\caption{The pipeline of the applied analyses}
\label{Figure:1}
\end{figure}

On completion of telling the pre-processing procedure for both raw and textual data, the process was attempted to summarize graphically for visualization in Figure~\ref{Figure:1}.

\subsection{Topic Models}

To effectively extract features from a large corpus of text data, different approaches have been introduced in \cite{egger_yu}. One such approach is topic modeling, which serves as the most commonly used technique. Topic modeling methods, which are known to be very successful in analyzing various terms in the corpus, are unsupervised machine learning methods \cite{9984466}. Although it is known that only statistical analysis is performed in the literature with the concerned topic modeling methods in this field, each model has its own unique characteristics \cite{egger_yu}. In this subsection, the unique features of the methods implemented in this study are presented. After highlighting the key aspects of topic modeling, detailed information on how the methodology is applied is provided.
\subsubsection{Latent Dirichlet Allocation}
\label{subsec:LDA}

LDA is a generative probabilistic model for discovering latent topics in a collection of documents \cite{jelodar}. It assumes that each document is a mixture of multiple topics, and each topic is characterized by a distribution over words \cite{blei}. The goal of LDA is to infer the underlying topic structure of the documents based on their word occurrences.

Variables used in LDA are defined below \cite{10.5555/944919.944937} :

\begin{itemize}
  \item \(D\): The set of documents in the corpus.
  \item \(K\): The number of topics to be discovered.
  \item \(V\): The vocabulary size (i.e., the number of unique words in the corpus).
  \item \(\alpha\): The hyperparameter for the Dirichlet prior to the per-document topic distributions.
  \item \(\beta\): The hyperparameter for the Dirichlet prior to the per-topic word distributions.
\end{itemize}

The generative process of LDA can be summarized as follows:

\begin{enumerate}
  \item For each document \(d\) in \(D\):
    \begin{itemize}
      \item Sample a distribution over topics from a Dirichlet distribution: \(\boldsymbol{\theta}_d \sim \text{Dirichlet}(\boldsymbol{\alpha})\).
    \end{itemize}
  \item For each word \(w\) in document \(d\):
    \begin{itemize}
      \item Sample a topic assignment from a multinomial distribution: \(z_{d,w} \sim \text{Multinomial}(\boldsymbol{\theta}_d)\).
      \item Sample a word from a multinomial distribution specific to the assigned topic: \(w_{d,w} \sim \text{Multinomial}(\boldsymbol{\phi}_{z_{d,w}})\).
    \end{itemize}
\end{enumerate}

In the generative process, \(\boldsymbol{\theta}_d\) represents the distribution of topics in document \(d\), and \(\boldsymbol{\phi}_{z_{d,w}}\) represents the distribution of words in topic \(z_{d,w}\).

The goal of LDA is to estimate the posterior distributions of the latent variables, given the observed words in the documents. This is done using Bayesian inference and the variational EM algorithm or Gibbs sampling.

The output of the LDA algorithm consists of two main components:

\begin{itemize}
  \item \(\boldsymbol{\theta}\): A matrix of shape \((D, K)\), where each row \(\boldsymbol{\theta}_d\) represents the topic distribution for document \(d\).
  \item \(\boldsymbol{\phi}\): A matrix of shape \((K, V)\), where each row \(\boldsymbol{\phi}_k\) represents the word distribution for topic \(k\).
\end{itemize}

By analyzing the inferred topic distributions in \(\boldsymbol{\theta}\) and word distributions in \(\boldsymbol{\phi}\), one can gain insights into the latent topics present in the corpus and their associated words.


Latent Dirichlet Allocation (LDA) has found extensive applications in various fields due to its ability to uncover latent topics within a collection of documents. LDA has been widely used in information retrieval, document clustering, topic modeling, sentiment analysis, and recommendation systems. Its advantages lie in its ability to handle large-scale document collections, automatically discover hidden thematic structures, and provide interpretability of topics through the distribution of words \cite{JEONG2019280}. LDA allows for unsupervised learning, enabling the exploration of data without prior knowledge of the underlying topics. One of the key properties of LDA is its generative nature, which allows for flexible modeling of documents as mixtures of topics \cite{egger_yu}. Additionally, LDA incorporates Dirichlet priors, which provide a way to control the sparsity and smoothness of the topic distributions. These properties make LDA a powerful tool for understanding and organizing large text corpora, facilitating information retrieval, and aiding in knowledge discovery.

\subsubsection{Non-negative Matrix Factorization}
\label{subsec:NMF}

Non-negative Matrix Factorization (NMF) is a dimensionality reduction technique that factorizes a non-negative matrix into two lower-rank non-negative matrices. It has been widely used in various applications such as image processing, text mining, and bioinformatics. NMF aims to discover latent patterns or topics present in the data, which can be represented as combinations of positive components. Variables used in NMF can be defined as follows:

\begin{itemize}
  \item \(X\): The input non-negative data matrix of size \(m \times n\).
  \item \(r\): The desired rank or the number of components in the factorization.
\end{itemize}

The goal of NMF is to find two non-negative matrices \(W\) and \(H\) of sizes \(m \times r\) and \(r \times n\), respectively, such that \(X \approx WH\). Here, \(W\) represents the basis matrix that captures the features or patterns in the data, and \(H\) represents the coefficient matrix that indicates the contribution of each basis to reconstruct the data \cite{egger2}.

The NMF algorithm can be summarized as follows:

\begin{enumerate}
  \item Initialize \(W\) and \(H\) with non-negative random values.
  \item Repeat until convergence:
    \begin{itemize}
      \item Update \(W\) by minimizing the reconstruction error: \(W \leftarrow W \odot \left(\frac{X}{WH}H^T\right)\), where \(\odot\) denotes element-wise multiplication.
      \item Update \(H\) by minimizing the reconstruction error: \(H \leftarrow H \odot \left(\frac{W^TX}{W^TWH}\right)\).
    \end{itemize}
\end{enumerate}

The algorithm iteratively updates \(W\) and \(H\) to minimize the difference between the input matrix \(X\) and its reconstruction \(WH\). This is often achieved by optimizing a suitable objective function, such as the Frobenius norm or Kullback-Leibler divergence. NMF possesses several properties that make it valuable in various applications. One important property is that the factorization is non-negative, which allows for parts-based representation and interpretation of the data. NMF can extract meaningful and sparse features, which is particularly useful for tasks like image analysis, where the data often exhibits a parts-based structure. Furthermore, NMF is a dimensionality reduction technique that can effectively reduce the dimensionality of high-dimensional data while preserving essential information.

NMF has been successfully applied in numerous domains, including text mining for topic modeling and document clustering, image processing for feature extraction and image segmentation, and bioinformatics for gene expression analysis and pattern discovery. Its ability to uncover latent patterns, handle non-negative data, and provide interpretable results has made NMF a popular and powerful tool in data analysis and machine learning \cite{dhillon}.


\subsubsection{Latent Semantic Analysis}
\label{subsec:LSA}
Latent Semantic Analysis (LSA) is a technique used for extracting and representing the latent semantic structure of a collection of documents \cite{barde}. It aims to capture the hidden relationships between terms and documents based on their co-occurrence patterns. LSA is commonly employed in information retrieval, text classification, and document clustering tasks. Variables used in LSA can be defined as follows:

\begin{itemize}
  \item \(D\): The set of documents in the corpus.
  \item \(V\): The vocabulary or set of unique terms in the corpus.
  \item \(X\): The term-document matrix of size \(|V| \times |D|\), where each entry \(X_{ij}\) represents the frequency or weight of term \(i\) in document \(j\).
\end{itemize}

The LSA algorithm can be summarized as follows:

\begin{enumerate}
  \item Construct the term-document matrix \(X\) based on term frequencies or other suitable weighting schemes.
  \item Perform Singular Value Decomposition (SVD) on \(X\) to obtain the factorization \(X = U\Sigma V^T\), where \(U\) is a \(|V| \times k\) matrix, \(\Sigma\) is a diagonal \(k \times k\) matrix, and \(V^T\) is a \(k \times |D|\) matrix. Here, \(k\) is the desired number of latent dimensions.
  \item Represent each document \(j\) as a low-dimensional vector by taking the corresponding row \(d_j\) of \(V\), i.e., \(d_j = V_{\cdot j}\).
  \item Compute the cosine similarity or other suitable distance measures between document vectors to perform various tasks like document retrieval, clustering, or classification.
\end{enumerate}

The SVD in LSA helps to identify the latent semantic dimensions that capture the underlying topics or concepts in the document collection. The resulting low-dimensional representations enable efficient computation of similarity measures between documents and the identification of related documents or topics.

LSA possesses several properties that make it useful in various applications \cite{dharmendra}. By reducing the dimensionality of the data, LSA helps to overcome the data sparsity problem and captures the latent semantic structure of documents. It can uncover hidden relationships and similarities between documents, even when they do not share many common terms. Moreover, LSA has been successfully applied in information retrieval systems to improve search accuracy, document clustering to group related documents, and text classification to categorize documents based on their topics or sentiments. It has also been used in tasks such as automated essay grading, recommender systems, and concept extraction from text corpora.

The LSA algorithm provides a powerful approach for exploring and understanding large document collections by revealing the underlying semantic structure. Its ability to handle synonymy, and polysemy, and capture latent relationships makes it a valuable tool in text analysis and information retrieval.

\section{Results and Discussions}
\label{sec:results}
This section presents the qualitative results of the research, focusing on the four key themes. Firstly, a comparison of the coherence scores of the described methods is presented taking various combinations of the text-related sets as the number of topics increases. In a second manner, consulting with the experts of the field the topics are expressed regarding the combination as well as the model which attained the highest coherence score. It is followed by looking at how consistent the determined topics are with the data re-collected in the literature. A further discussion is conducted by studying the new data to extract the emerging topics for shedding new light on the field for the researchers. To avoid confusion, it is vital to emphasize at the beginning of the section that
\begin{itemize}
\item All analyses to determine both the number of topics and the combination of features have been conducted on the data from 01 January 2008 to 31 December 2021.
\item The validation has been provided on the data collected from 01 January 2008 to 24 April 2023.
\item The last part of the study for predicting upcoming trends has been conducted on data from 01 January 2008 to 24 April 2023.
\end{itemize}

To accomplish the first scope, data from 01 January 2008 to 31 December 2021 has been considered. Different combinations of sets consisting of "Abstract", "Title", and "Keywords" (SET 1), "Title" and "Web of Science Categories" (SET 2), "Abstract" and "Keywords" (SET 3), and the union of "Title" and "Keywords" (SET 4) have been generated. All the selected models, namely LDA, NMF, and LSA, are performed on the specified sets by varying the number of topics. The coherence scores of the models have been presented in Table~\ref{tab:1}.

\begin{figure}[h!]
\centering
\includegraphics[width=0.8\textwidth]{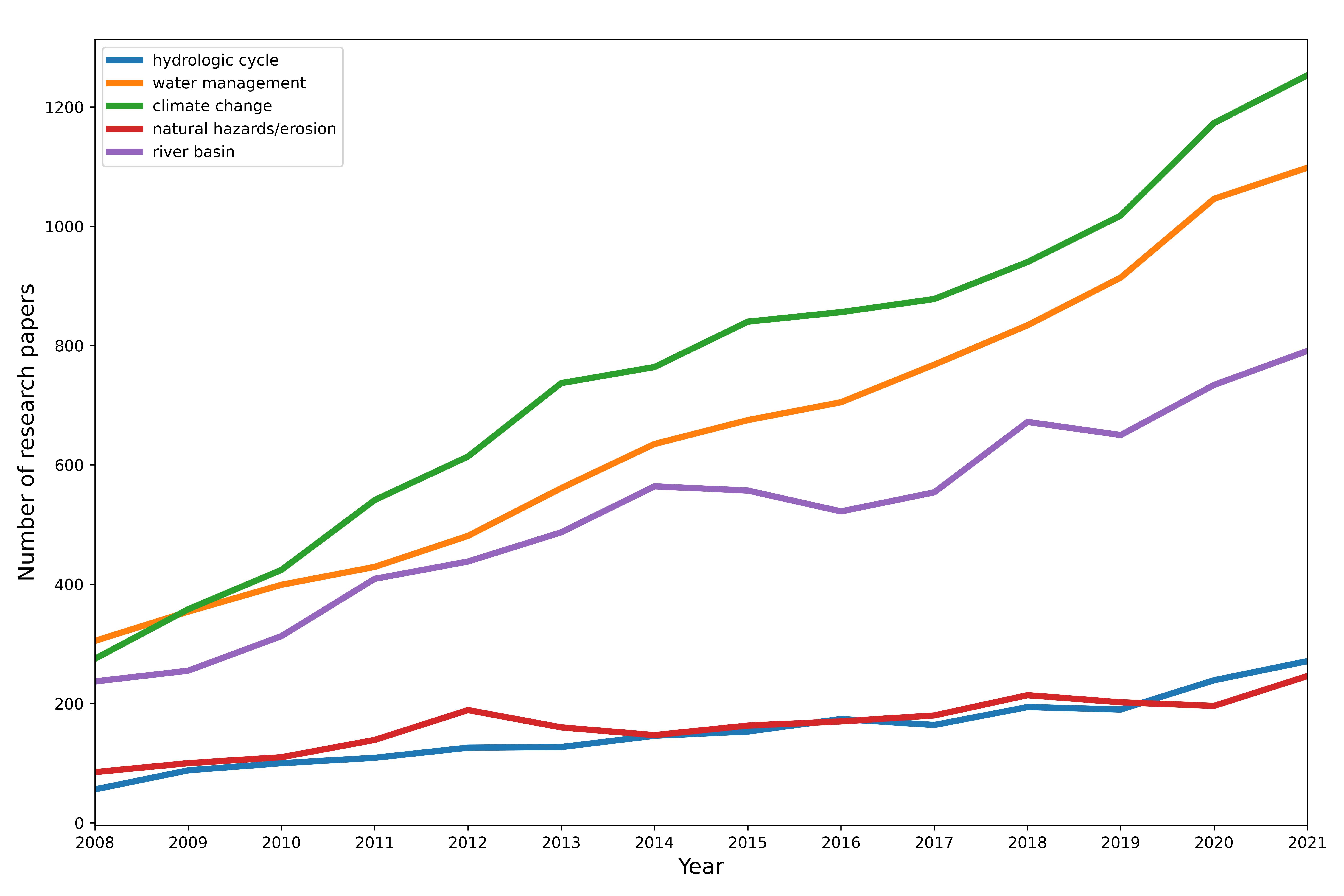}
\caption{The behavior of the number of articles related to attractive topics determined by the field experts over the years. For the depicted figure, the time interval has been selected from 01 January 2008 to 31 December 2021. The data belonging to 2022 has been ignored without losing generality.}
\label{Figure:2}
\end{figure}

\begin{table}[th]
\caption{\label{tab:1}Coherence score of the models according to the number of topics}
\centering%
\begin{tabular}{cllll}
\hline
$\begin{array}{c}
\text{Number\,of} \\
\text{Topics}
\end{array}$  & Sets  & LDA & NMF & LSA \\ \hline
& SET 1 & $0.5463$ & $0.5835$ & $0.5586$ \\
5& SET 2& $\mathbf{0.5917}$ & $0.6269$ & $\mathbf{0.6153}$\\
&SET 3& $0.5659$ & $0.5924$ & $0.5560$\\
&SET 4 & $0.5081$ &  \boxit{1.2cm}$\mathbf{0.6521}$ & $0.5818$\\
\hline
\hline
& SET 1 & $0.5034$ & $0.4800$ & $0.4570$\\
10& SET 2 & $0.4757$ & $0.4926$ & $0.4152$\\
&SET 3& $0.5095$ & $0.4889$ & $0.4672$\\
&SET 4& $0.4770$ & $0.5438$ & $0.4617$\\
\hline
\hline
&SET 1& $0.5001$ & $0.4628$ & $0.4061$\\
15&SET 2 & $0.3796$ & $0.4540$ & $0.3545$\\
&SET 3& $0.4912$ & $0.4687$ & $0.4066$\\
&SET 4& $0.4380$ & $0.5042$ & $0.3867$\\
\hline
\end{tabular}%
\end{table}
What stands out from Table~\ref{tab:1} is the superiority of the NMF method. The highest coherence score has been recorded as 0.6521 with the NMF method in the combination of "Title" and "Keywords" for 5 topics. The topics identified in this response have been analyzed in Table~\ref{tab:2}.
\begin{table}[h!]
\caption{\label{tab:2} Topics obtained by NMF in line with the experts of the field}
\begin{tabular}{ p{0.6cm} p{2.2cm} p{4cm} p{4cm}  }
\hline
   & \multicolumn{3}{c}{\textbf{NMF}}  \\
\cline{2-4}
No.& \textbf{MTF*}  & \textbf{Full content*} & \textbf{Keywords*}\\
\cline{1-4}
 1.  &  hydrologic cycle & Simulating a hydrologic cycle model with rainfall, base flow, groundwater flow, and runoff by taking uncertainty into account. & model, use, rainfal, simul, data, hydrolog, runoff, base, groundwat, uncertainti \\
 2. & water management &  Managing water resources quality with controlling the water balance of groundwater and surface water such as lakes etc. with isotopograhic imaging.  &  water, groundwat, manag, qualiti, resourc, surfac, isotop, balanc, lake, use \\
  3. &climate change  &  Determining climate change impact in regions, which have river basin, by precipitation variability on lands.  &  climat, chang, impact, land, basin, river, use, variabl, precipit, region \\
4. & natural hazards/ erosion &  Sensitivity analysis of erosion on land surface with using moisture, vegetation ratio, carbon ratio in soil.  &   soil, moistur, land, surfac, use, data, eros, sens, carbon, veget \\
5. & river basin & Analysis of flood risk, sedimentation, and stream of a river basin via hydrological flows for wetland conditions.  &  hydrolog, river, basin, flow, flood, analysi, sediment, wetland, catchment, stream\\
 \hline
\end{tabular}
    \begin{tablenotes}\footnotesize
      \item Full content: the possible match subjects.
      \item MTF: The main topic of the Field
      \item Keywords: The results of the NMF method
    \end{tablenotes}
\end{table}
Table~\ref{tab:2} also provides the possible topics of the field consulting the field experts. Additionally, Figure~\ref{Figure:2} indicates the effects of topics determined by the field experts over the years. The attained results are recorded considering the research items of the trained data from 01 January 2008 to 31 December 2021. It is worth noting that the counts of research items in the latest year, 2022, have been excluded in Figure~\ref{Figure:2} due to the time spans covering only January and March in that year.
Figure~\ref{Figure:2} highlights that the number of research related to topics of "climate change", "river basin", and "water management" were superior to those of "natural hazards/erosion" and "hydrologic cycle". Moreover, Figure~\ref{Figure:2} also indicates that the greatest acceleration belongs to "water management" between 2016 and 2019. Furthermore, the growth of the studies on "climate change" is greater than those of the other topics. One can state that the subjects that are shown in Figure~\ref{Figure:2} are directly related to the cause-and-effect relationship. More explicitly, due to awareness of climate change impacts on water balance in the recent couple of decades, studying climate change has become a major subject in hydrology. It is clear that climate change affects all hydrologic events and water balance deficiency as a result. As is seen in Figure~\ref{Figure:2}, the number of published papers on climate change and water management intersected around 2009 and kept the gap in the same range in the following years. This is because most of the water management papers are related to the reason for the problem of "climate change" during that period. On the other hand, the "river basin” subject is also observed as the same increasing trend as "climate change” and "water management" topics. This result is very much expected because the primary water source of human beings is surface water like rivers. "climate change" has a direct impact on river basins such as flow rate, sediment transportation, etc.  As a result of the change river characteristic reflects the adjustment of water balance as well. Therefore, it can easily be assumed direct relativeness between these subjects. Moreover, the change in the river’s stream can cause flooding, which is one of the reasons for erosion. Due to fast-passing fresh water in the basin, vegetation loss may occur, catchment topography may be altered, surface water balance may deteriorate, groundwater level may be depleted, etc. These natural events are followed by changes in the hydrologic cycle.

The validation of the obtained result has been tested on a new data set between 01 January 2022 and 24 April 2023 to see how effective the model is.  After extracting from WOS on the specified time span and conducting a similar pre-process procedure, further manipulation has been done which removes the duplications compared to the trained if exist. Once the data has been prepared, the determined topics have been discussed. Table\ref{tab:3} describes the percentages of how many research items have been studied in almost one year period with the related topics.
\begin{table}[ht]
\begin{minipage}[b]{0.45\linewidth}
\centering
\includegraphics[width=60mm]{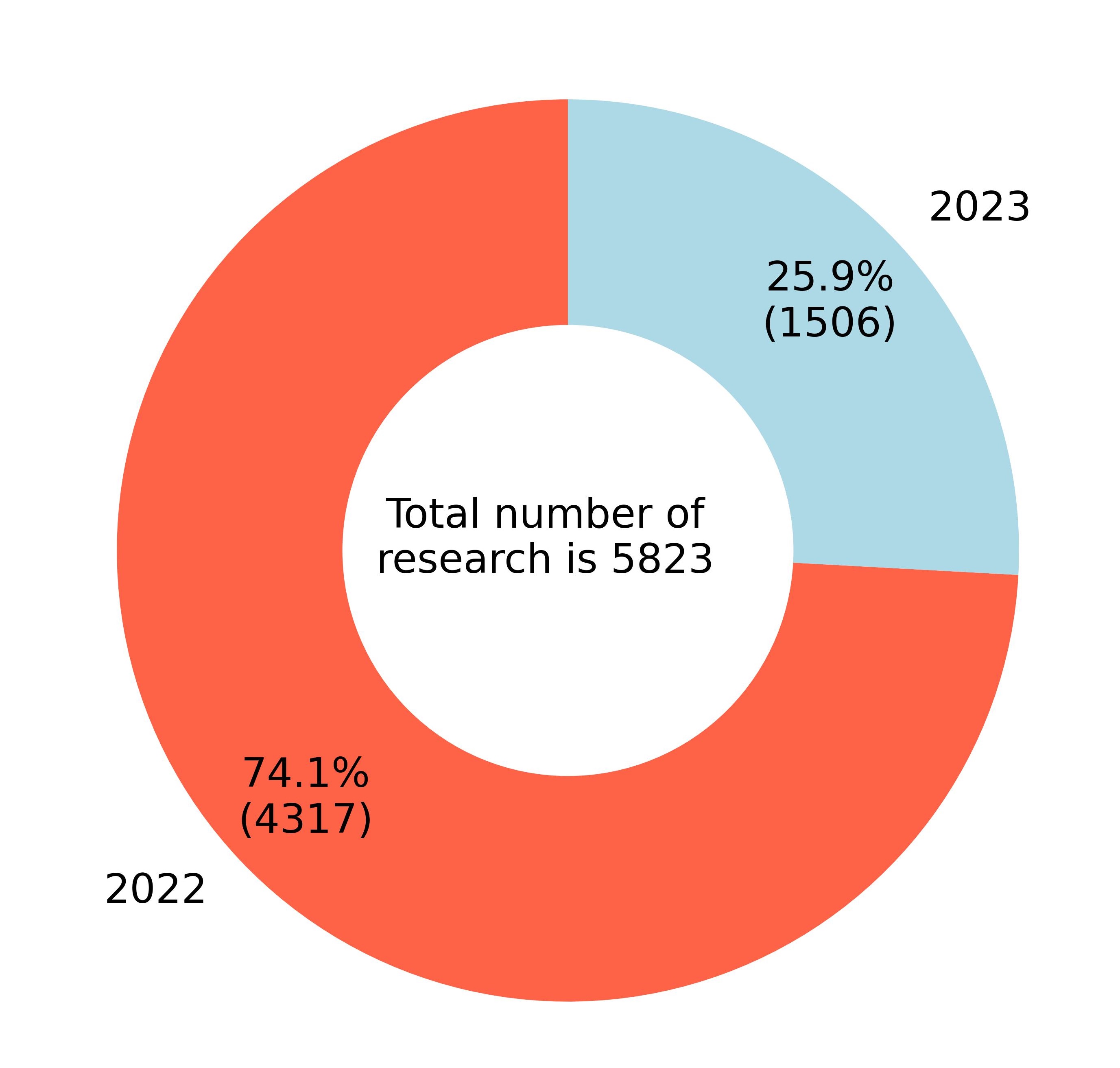}
\captionof{figure}{The number of new research items on the\\ validation set, that is the data between\\ 01 January 2022 and 24 April 2023}\label{figure:3}
\end{minipage}
\hfill
\begin{minipage}[b]{0.45\linewidth}
\begin{center}
\begin{tabular}{l c }
  \hline
  \textbf{Topics}           & \textbf{\% Research Items} \\ \hline
  hydrologic cycle & 4.91   \\
  water management & 21.16  \\
  climate change   & 24.85  \\
  erosion          & 4.59  \\
  river basin      & 15.89  \\
  \hline
\end{tabular}
\caption{The percentages of the new research items in one year\\(on validation set)}\label{tab:3}
\end{center}
\end{minipage}
\end{table}
It is apparent from Table~\ref{tab:3} that the studies on the topics of "climate change", "river basin", and "water management" still preserve their influence in the field of hydrology.

\begin{figure}[h!]
\centering
\includegraphics[width=0.8\textwidth]{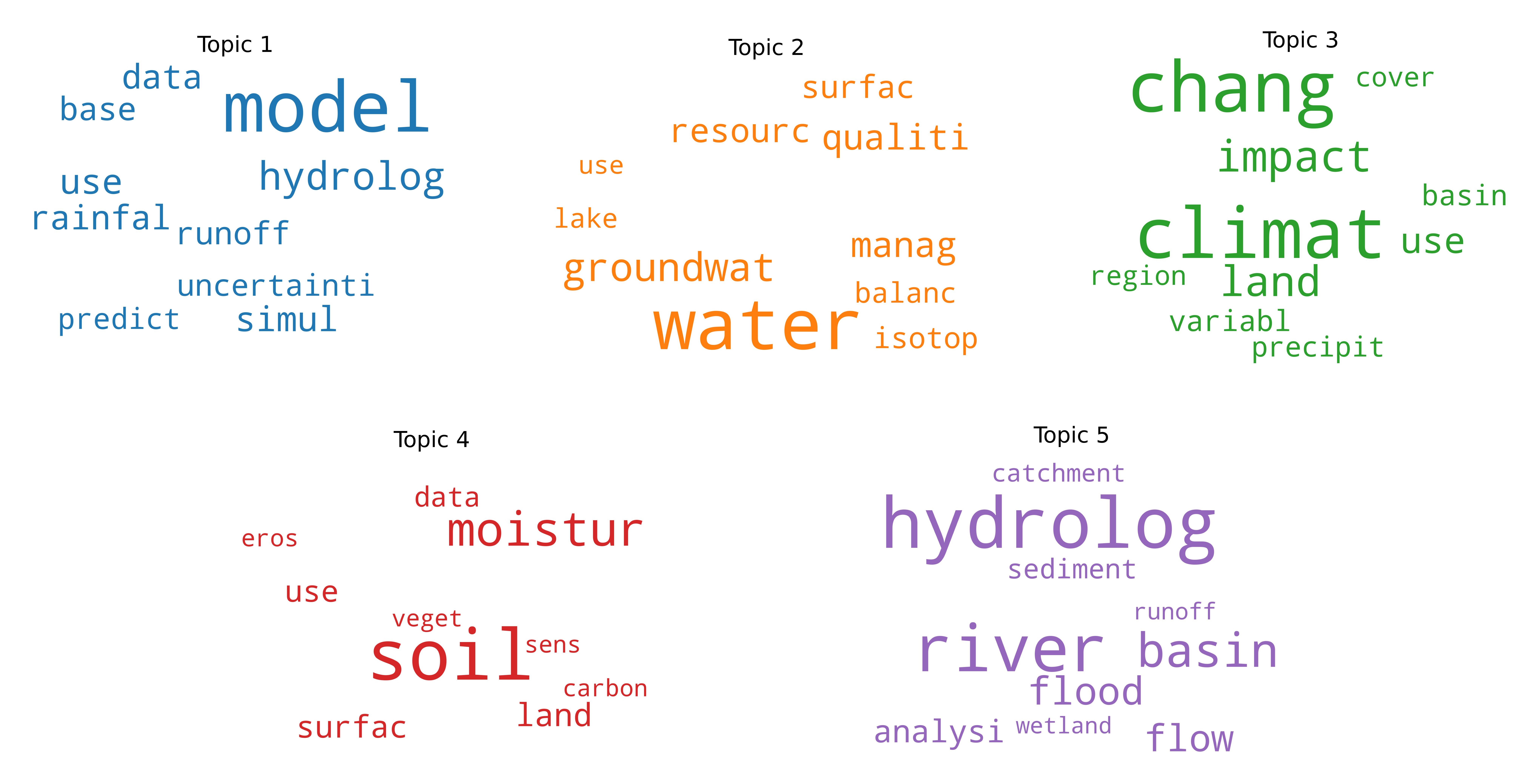}
\caption{Forecasted topics of the research articles published from 01 January 2008 to 24 April 2023 using NMF with 0.5037 coherence score}
\label{Figure:4}
\end{figure}

Before ending the section, a further analysis has been conducted for future projection as our final target. To do so, extended data composed of the research items published between 01 January 2008 and 24 April 2023 has been constructed by collecting the data from WOS and conducting similar preprocessing to those of the trained data. Thereafter, the NMF method has been studied once more on the extended data by selecting 5 topics on SET 4 ("Title" and "Keywords"). The topics obtained from the experiment within the 0.5037 are illustrated in Figure~\ref{Figure:4}. These results are in accord with previous analysis indicating that the impacts of "climate change", "river basin", "water management", "natural hazards/erosion", and "hydrologic cycle" topics remain attractive.

\section{Conclusion}
\label{sec:con}
Identifying trends can be significant in addressing academics, scientists, and organizations in the new global age. However, tracking of emerging topics individually suffers from abundant literature. Thus, any research done to suggest the trend themes in an area to anticipate future issues is important. The essential purpose is to present trend topics to the interested reader in the field of "Hydrology". To do so, two different periods have been examined. Firstly, 01 January 2008 to 31 December 2021 is used for extracting the topics of 2022. After validating the obtained trends, the data from 01 January 2008 to the end of April 2023 has been used to forecast future perspectives. In both data sets, the data preparation has been accomplished via three strategies: a gathering of data, preprocessing of the data and feature extraction of the bibliographic items of research, and identifying trend topics. Based on the purpose of the study, several topic models consisting of Latent Dirichlet Allocation (LDA), Non-negative Matrix Factorization (NMF), and Latent Semantic Analysis (LSA) have been taken into account for comparison. Besides the number of topics, another comparative work has been conducted, simultaneously, on the combination of features such as "Abstract", "Title", and "Keywords"; "Title" and "Web of Science Categories"; "Abstract" and "Keywords"; and the union of "Title" and "Keywords". The highest coherence score has been attained in the combination of "Title" and "Keywords" for 5 topics via the NMF method ($C_V=0.6521$). The observed trend topics corresponding to the best coherence score have been presented and assessed by the field. In line with the opinions of the field, the attractive topics of 2022 in Hydrology have been recorded as follows: climate change", "river basin", "water management", "natural hazards/erosion", and "hydrologic cycle". Further analyses have been conducted to verify the plausibility of the results in the upcoming year, 2023, and to predict the projection of future trends. The results have also confirmed the popularity of these topics in 2023. Furthermore, it is worth noting that these topics keep their effect on the field in 2023. From the hydrological point of view, the results can be interpreted as follows: direct relativeness between these subjects can be simply inferred. The flow rate, transit of sediment, and other factors are directly impacted by "climate change" on river basins. Since river characteristics have changed, the water balance has also been altered. Additionally, flooding may result from the river's stream change, which is one of the causes of erosion. The hydrologic cycle alters after certain natural occurrences such as fast-moving fresh water in the basin may cause a variety of negative effects, including plant loss, topographic changes in the catchment, deterioration of the surface water balance, depletion of groundwater levels, etc. Taken together, this study is a candidate to be a pioneer of an exploratory and interpretative study in the field.

\section*{Acknowledgments}

This work was supported by the Center of Scientific Research Projects of the Izmir Katip Celebi University [Grant Number: 2022-GAP-MUMF-0029]. The authors would like to present their gratefulness to the Izmir Katip Celebi University for supporting the project. 

\section*{Declarations}
The authors declare that they have no known competing financial or personal interests that could have seemed to affect the work reported in this study.

\bibliographystyle{plain}

\bibliography{main}







\end{document}